\documentclass[prl,twocolumn,superscriptaddress,preprintnumbers,amsmath,amssymb,aps]{revtex4}
\usepackage{graphicx}
\usepackage{amsmath}
\usepackage[dvipsnames]{color}
\usepackage{bm}

\usepackage[colorlinks=true,citecolor=blue]{hyperref}
\hypersetup{colorlinks=true,citecolor=blue,linkcolor=red,urlcolor=blue}

\begin{document}

\sloppy

\title{Cavity exciton-polaritons in two-dimensional semiconductors from first principles}

\author{Dino Novko}
\affiliation{Institute of Physics, Bijeni\v cka 46, 10000 Zagreb, Croatia}
\affiliation{Donostia International Physics Center (DIPC), P. Manuel de Lardizabal, 4, 20018 San Sebastián, Spain}
\author{Vito Despoja}
\email{vdespoja@ifs.hr}
\affiliation{Institute of Physics, Bijeni\v cka 46, 10000 Zagreb, Croatia}
\affiliation{Donostia International Physics Center (DIPC), P. Manuel de Lardizabal, 4, 20018 San Sebastián, Spain}

\begin{abstract}
Two-dimensional (2D) semiconducting microcavity, where exciton-polaritons can be formed, constitues a promising setup for exploring and manipulating various regimes of light-matter interaction. Here, the coupling between 2D excitons and metallic cavity photons is studied by using first-principles propagator technique. The strength of exciton-photon coupling is characterised by its Rabi splitting to two exciton-polaritons, which can be tuned by cavity thickness. Maximum splitting of 128\,meV is achieved in phosporene cavity, while remarkable value of about $440$\,meV is predicted in monolayer hBN device. The obtained Rabi splittings in  WS$_2$ microcavity are in excellent agreement with the recent experiments. Present methodology can aid in predicting and proposing potential setups for trapping robust 2D exciton-polariton condensates.
\end{abstract}

\maketitle

The interplay between cavity photons and matter can result in formation of hybrid polarization-photon modes, commonly called polaritons\,\cite{polaritons}. Due to their dual light-matter nature, these bosonic quasiparticles are enriched with expectional physical properties, such as small efective mass and nonlinearity, absent in the matter outside the optical cavity. As such, polaritons are expected to show exotic physical phenomena like light-induced superconductivity\,\cite{super1,super2}, Bose-Einstein condensation\,\cite{bec1,bec2,bec3}, polariton superfluditiy\,\cite{superfluid}, and quantized vortices\,\cite{vortex}. Ideal platforms for studying strong light-matter interactions are gapped systems, such as semincondutors\,\cite{bec3,1st-qwell} or molecules\,\cite{molecule1}, placed in microcavity devices, where exciton-polaritons are formed. Cavity exciton-polaritons, showing a plethora of quantum effects, are routinely observed in bulk\,\cite{GaN1,GaN2,Peroskite_ACSPhotonics,ex-po-perovskite} and quantum well systems\,\cite{1st-qwell,GaAsQW}, e.g., devised from GaAs\,\cite{1st-qwell}.

Two-dimensional (2D) materials, such as semiconducting monolayers, thin heterostructures and films, are even more attractive than their bulk counterparts, due to the reduced Coulomb screening and the corresponding large exciton binding energies\,\cite{TMD-exc1,TMD-exc2,TMD-exc3,TMD-exc4,hBN-exc,ph-ex1,ph-ex2,ph-ex-EXP2} that enable formation of well-defined exciton-polaritons even at room temperatures\,\cite{Nature_Polaritons}. The first 2D exciton-polariton were realized in monolayer of transition metal dichalcogenide (TMD) MoS$_2$, where Rabi splitting between exciton and cavity photon of $\sim 50$\,meV was observed\,\cite{ex-pol1}. Further photoluminescence studies showed clear anticrossing behaviour and splitting of exciton-polariton in other 2D TMD cavity devices, e.g., in MoSe$_2$\,\cite{ex-pol22}, WS$_2$\,\cite{ex-pol-WS2}, WSe$_2$\,\cite{ex-pol2,ex-po-WSe2}, and in MoSe$_2$-WSe$_2$ heterostructure\,\cite{Cavity_interlayer}. In addition, real-space imaging of exciton-polaritons was done by means of near-field scanning optical microscopy for WSe$_2$ thin films\,\cite{waveguide}. These TMD semincondutor microcavities are especially appealing due to their charge tunnability, coupled spin and valley degrees of freedom, as well as ability to form heterostructures, and are thus able to display valley-polarized exciton-polaritons\,\cite{valleypol}, polaron-polaritons\,\cite{polaronpol}, and interlayer exciton-polaritons\,\cite{Cavity_interlayer}.

Despite this enourmous interest in exciton-polaritons and seminconductor microcavity devices, a complementary microscopic theories that are able to scrutinize the cavity photon-exciton coupling on the quantitative and predictive level are still rare. In addition, the majority of the microscopic descriptions are based on simple model Hamiltonians describing exciton-photon interactions in microcavity\,\cite{Theo-ex-pol0,Theo-ex-pol1,Theo-ex-pol3,Theo-ex-pol2,Theo-ex-pol4,Theo-ex-pol5}. Recently, a more rigorous \emph{ab initio} theoretical description of exciton-polaritons in TMD microcavity was provided in the framework of the quantum-electrodynamical Bethe-Salpeter equation\,\cite{Rubio_CAVITY-TMD}, where the excitons are calculated from first-principles Bethe-Salpeter equation, and the electromagnetic field is described by quantized photons. However, the coupling between excitons and photons is left to be arbitrary. This study showed how excitonic optical activity and energetic ordering can be controlled via cavity size, light-matter coupling
strength, and dielectric environment.

Here, we present a fully quantiative theory of 2D exciton-polaritons embedded in plasmonic microcavity that is able to analyze and predict light-matter coupling strengths for various cavity settings. We study cavity exciton-polaritons in three prototypical two-dimensional single-layer semiconductors, i.e., single-layer black phosphorus or phosphorene (P$_4$), WS$_2$, and hexagonal boron nitride hBN. The results show a clear Rabi splitting between 2D exciton and cavity photon modes as well as high degree of tunability of Rabi (light-matter) coupling $\Omega$ as a function of microcavity thickness. In the case of WS$_2$, for the usual experimental setup where cavity size is around $d\sim 1\,\mu m$ we obtain splittings of $\Omega\sim 42$\,meV and $\Omega\sim 64$\,meV for principal and second cavity modes, in line with the experiments\,\cite{ex-pol1,ex-pol-WS2}. Also, in all cases larger coupling strengths $\Omega$ are found for larger photon confinements when $d<1\,\mu m$ as well as for higher cavity modes $n$. Interestingly, the ultraviolet (UV) exciton in hBN shows a very strong exciton-cavity photon coupling of about $\sim440$\,meV and a possiblity of Bose-Einstein condensation.

\begin{figure}[!t]
\centering
\includegraphics[width=8.5cm,height=4.2cm]{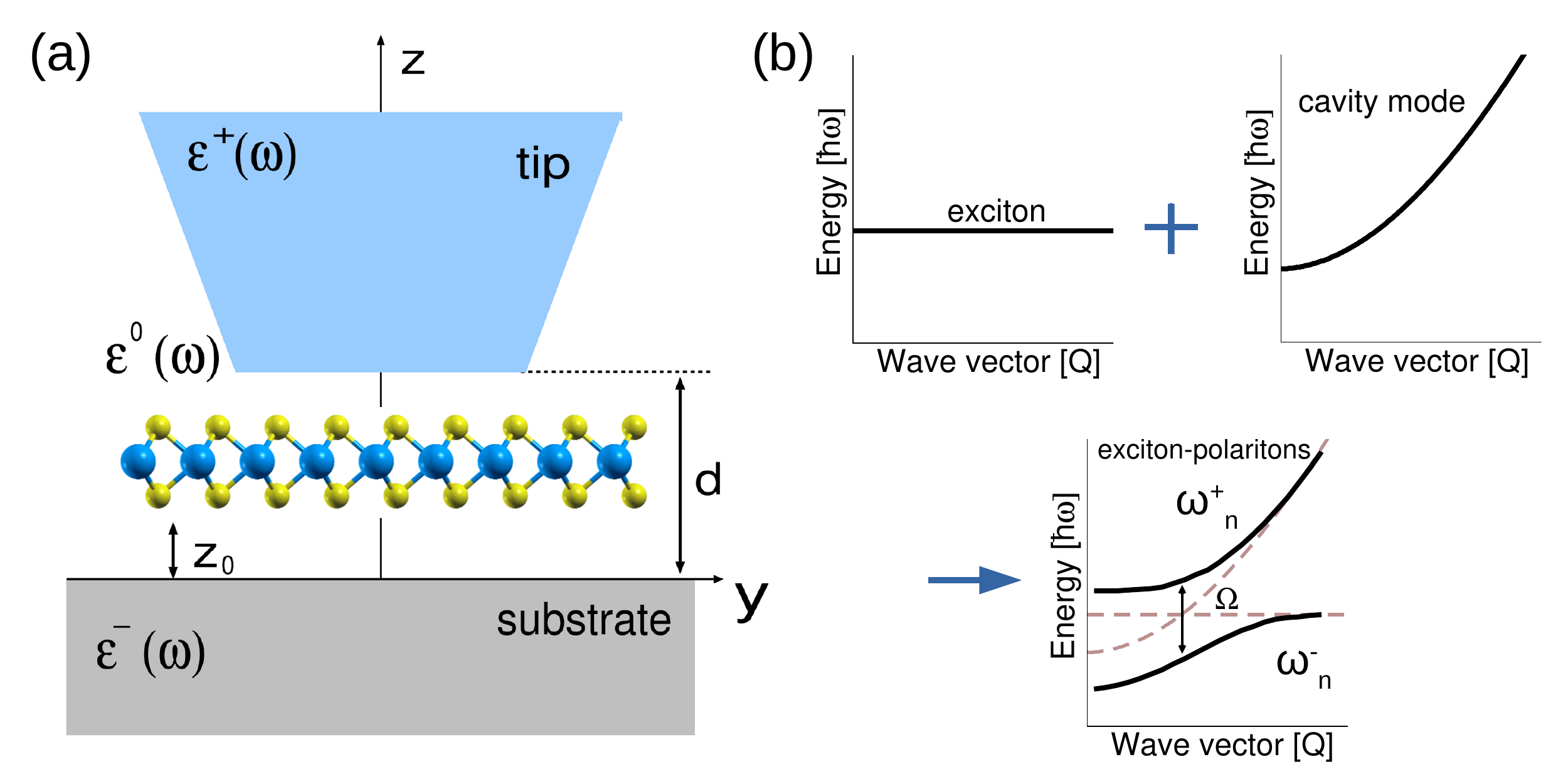}
\caption{(a) The schematic of a microcavity device. The 2D crystals described by optical conductivity $\sigma_\mu(\omega)$ is inserted in microcavity which consists of substrate, dielectric media and tip described by macroscopic dielectric functions $\epsilon^-(\omega)$, $\epsilon^0(\omega)$  and 
$\epsilon^+(\omega)$, respectively. (b) The intercation between 2D exciton and cavity mode results in creation 
of exciton-polaritons. The coupling strength measure is Rabi splitting $\Omega$.}
\label{Fig1}
\end{figure}

In this work both excitons and photons are described by bosonic propagators $\sigma$ and 
$\Gamma$, repectively,  which are derived from first principles. The 2D crystal optical 
conductivity $\sigma$ is calculated using {\em ab initio} RPA+ladder method\,\cite{ExPol-theory} , and propagator of cavity photons  
$\Gamma$ is derived by solving the Maxwell's equations for planar cavity decribed by local dielectric 
function $\epsilon$ (see Supplemental Material\,\cite{sm}).  
The exciton-photon coupling is achieved by dressing the cavity-photon propagator $\Gamma$ with excitons at the RPA level.
Thus obtained results are therefore directly comparable with the experiments. As illustrated in Fig.\,\ref{Fig1}(a) the microcavity device consists of substrate, tip and dielectric media in between, described by local dielectric functions $\epsilon^+$, $\epsilon^-$ and $\epsilon^0$, respctively. The 2D semiconducting crystal defined by optical conductivity $\sigma_{\mu}(\omega)$ is immersed in a dielectric media at a height $z_0$ relative to the substrate. The substrate occupy region $z<0$, tip occupy region $z>d$, and dielectric media  occupi region $0<z<d$.
In such semiconducting microcavity setup the coupling between the exciton and cavity photon is expected to result in the splitting of exciton-polariton to the lower and upper polariton branches (LPB and UPB), which we shall refer as $\omega^-_n$ and $\omega^+_n$, respectively [see Fig.\,\ref{Fig1}(b)]. 

\begin{figure}[!t]
\centering
\includegraphics[width=7.3cm,height=5.0cm]{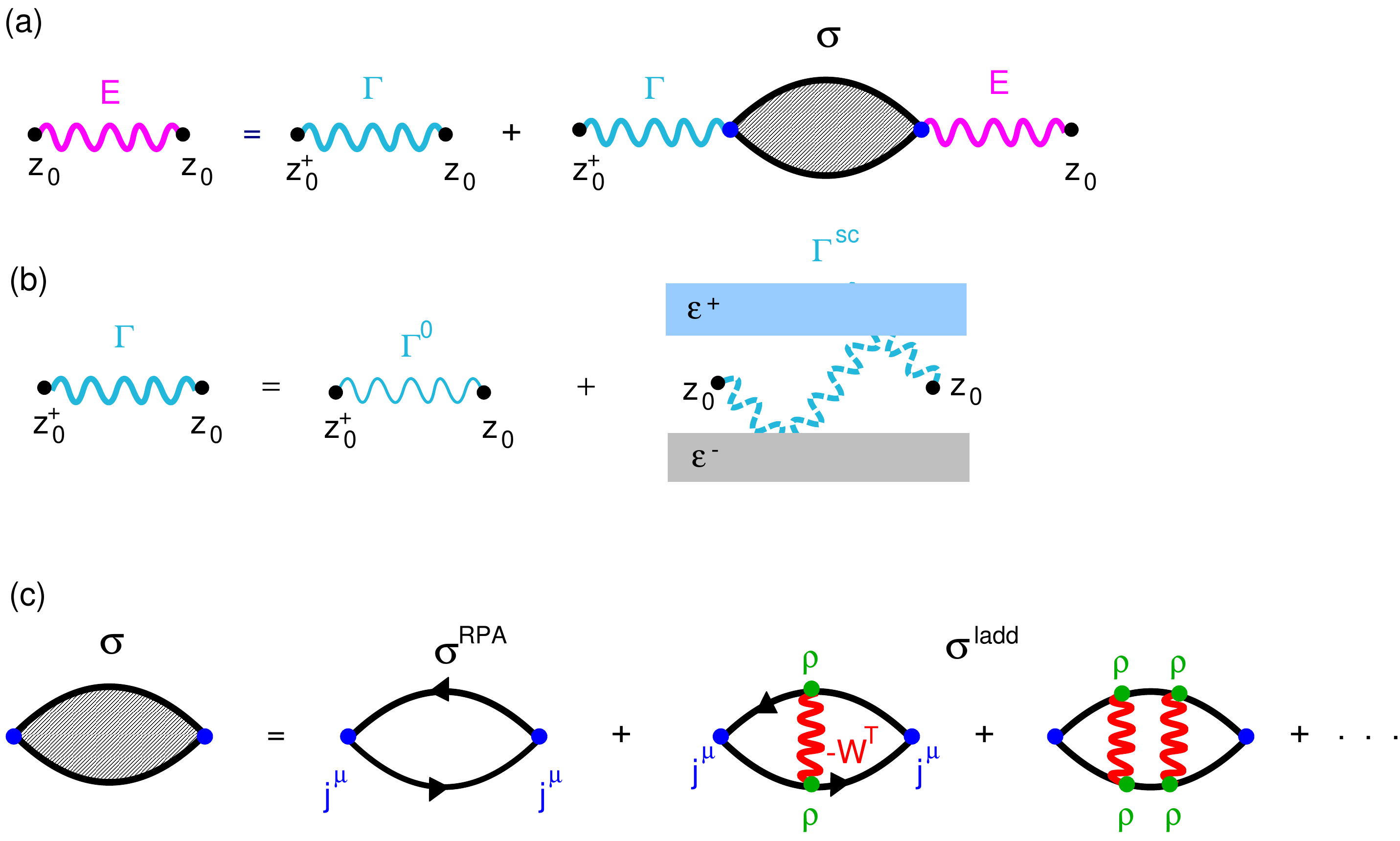}
\caption{(a) Dyson's equation for the propagator of the electrical field ${\cal E}$ in a microcavity device. (b) Propagator of the electrical field $\Gamma=\Gamma^0+\Gamma^{sc}$, in absence of 2D crystal (i.e., when $\sigma_\mu=0$).  
(c) Perturbative expansion of optical conductivity $\sigma_\mu(\omega)$. 
Blue dots represent the current vertices $j^{\mu}$,  green dots represent the charge vertices $\rho$ and red wavy lines represent the screened Coulomb interaction $W^T$.}
\label{Fig2}
\end{figure}

The quantity from which we extract the information about the 
electromagnatic modes in microcavity setup is electrical field propagator ${\cal E}_{\mu\nu}$ which, by 
definition \cite{Tomas}, propagates the electrical field produced by point oscillating dipole 
${\bf p}_0e^{-i\omega t}$, i.e. ${\bf E}(\omega)={\cal E}(\omega){\bf p}^0$. Assuming that the 2D crystal, substrate and tip satisfy planar 
symmetry (in $x-y$ plane) the propagator ${\cal E}$  in $z=z_0$ plane satisfies matrix 
equation 
\[{\bf{\cal E}}({\bf Q},\omega)={\bf\Gamma}({\bf Q},\omega)+
{\bf\Gamma}({\bf Q},\omega){\bf \sigma}(\omega){\bf {\cal E}}({\bf Q},\omega)
\]
as illustrated by Feynmans diagrams in Fig.\,\ref{Fig2}(a) (see also Sec.\,S1.A in Ref.\,\cite{sm}).
Here ${\bf \Gamma}={\bf \Gamma}^0+{\bf\Gamma}^{sc}$ represents the propagator of electrical field, in absence of 2D crystal, ie. when  $\sigma=0$. 
The propagator ${\bf \Gamma}^0$ represents the ``free'' electrical field and the propagator of scattered 
electrical field ${\bf\Gamma}^{sc}$ results in multiple reflections at the microcavity interfaces, as ilustrated in Fig.\,\ref{Fig2}(b). In order to simplify the interpretation of the results we suppose that the dielectric media 
is vacuum ($\epsilon^0=1$), and we suppose that tip and substrate are made of the same material ($\epsilon^-=\epsilon^+=\epsilon$). 
In order to support well-defined cavity modes, these materials should be highly reflective in the exciton frequency region 
$\omega\approx \omega_{ex}$, which is satisfied if $\omega_{ex}<\omega_p$, where $\omega_p$ 
is the bulk plasmon frequency. For the P$_4$ and WS$_2$ monolayers where exciton energies are $\hbar\omega_{ex}<3.0$\,eV, we chose that the substrate and tip are made of 
silver ($\omega_p\approx 3.6$\,eV). One the other hand, for the single-layer hBN where exciton energy is $\hbar\omega_{ex}=5.67$\,eV we chose aluminium  
($\omega_p\approx 15$\,eV). Both silver and  aluminium macroscopic dielectric 
functions $\epsilon(\omega)$ are determined as well from the first principles (see Ref.\,\cite{sm}).
As an example, the intensity of electromagnatic modes ($-{\rm Im}\Gamma_x$) in silver microcavity is shown in Fig.\,S2 in Ref.\,\cite{sm}. 
Figure \ref{Fig2}(c) shows the perturbative expansion of optical conductivity 
$\sigma_{\mu}(\omega)=\sigma^{RPA}_{\mu}(\omega)\ +\sigma^{ladd}_{\mu}(\omega)$,
\begin{figure*}[!t]
\includegraphics[width=4cm,height=4cm]{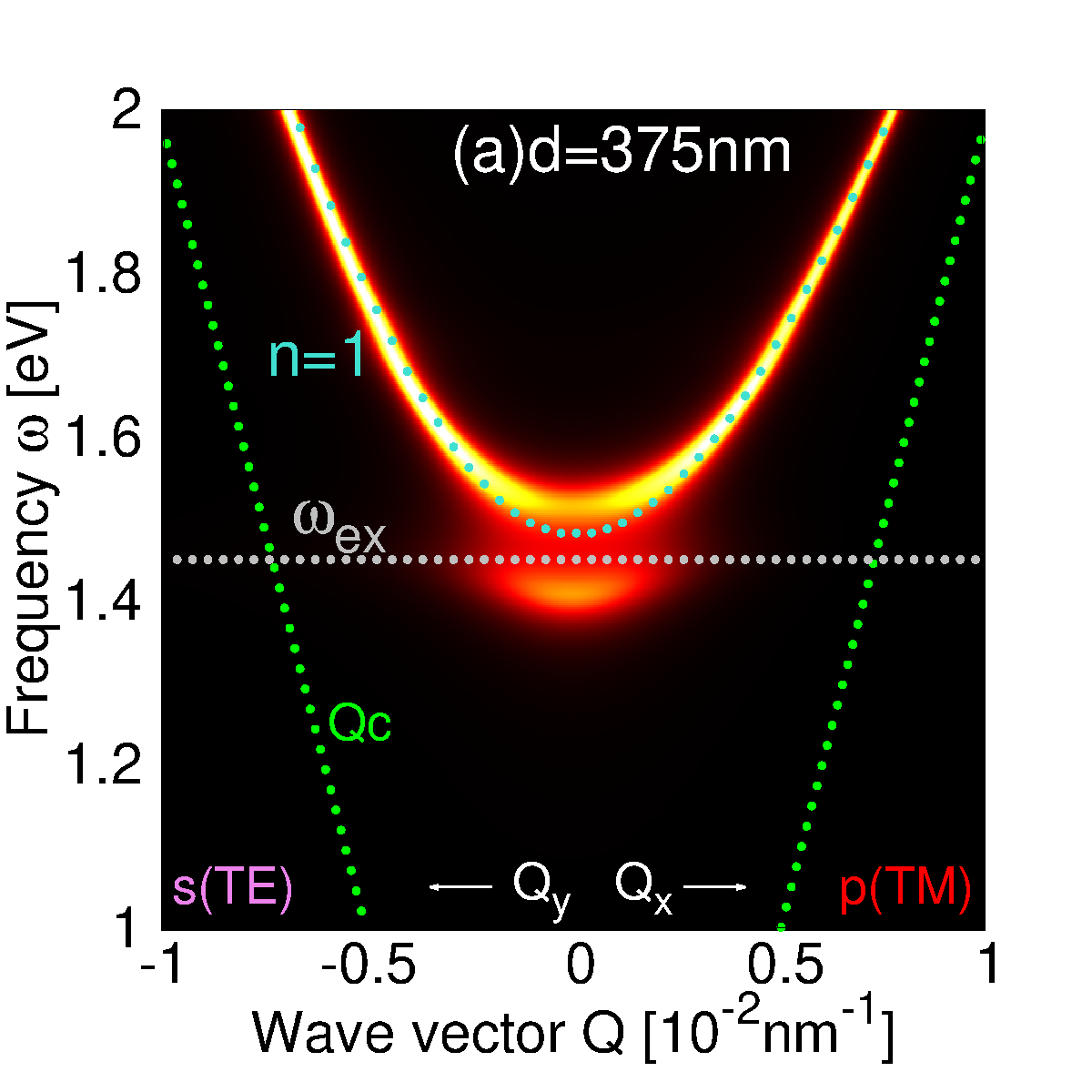}
\includegraphics[width=4cm,height=4cm]{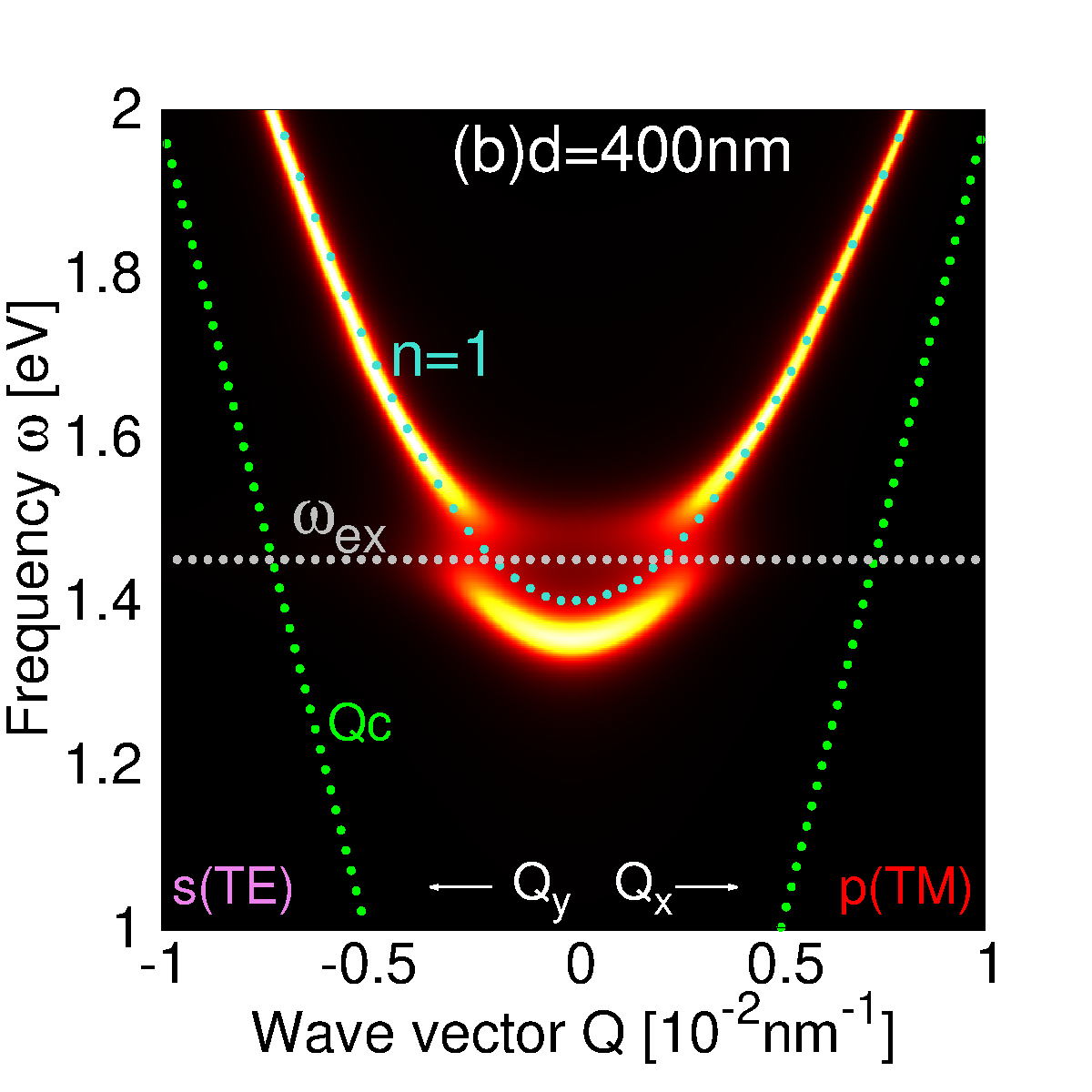}
\includegraphics[width=4cm,height=4cm]{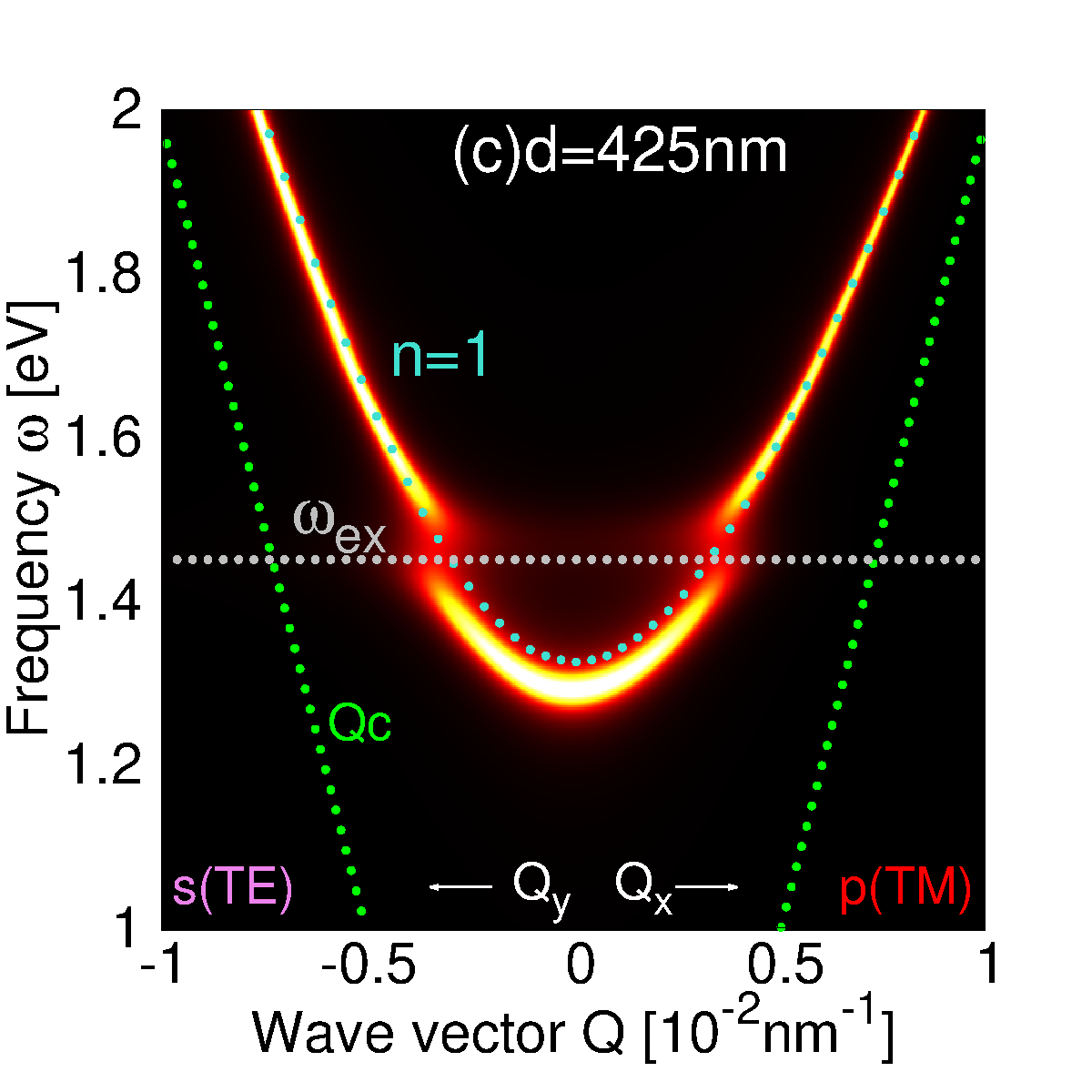}
\includegraphics[width=4cm,height=4cm]{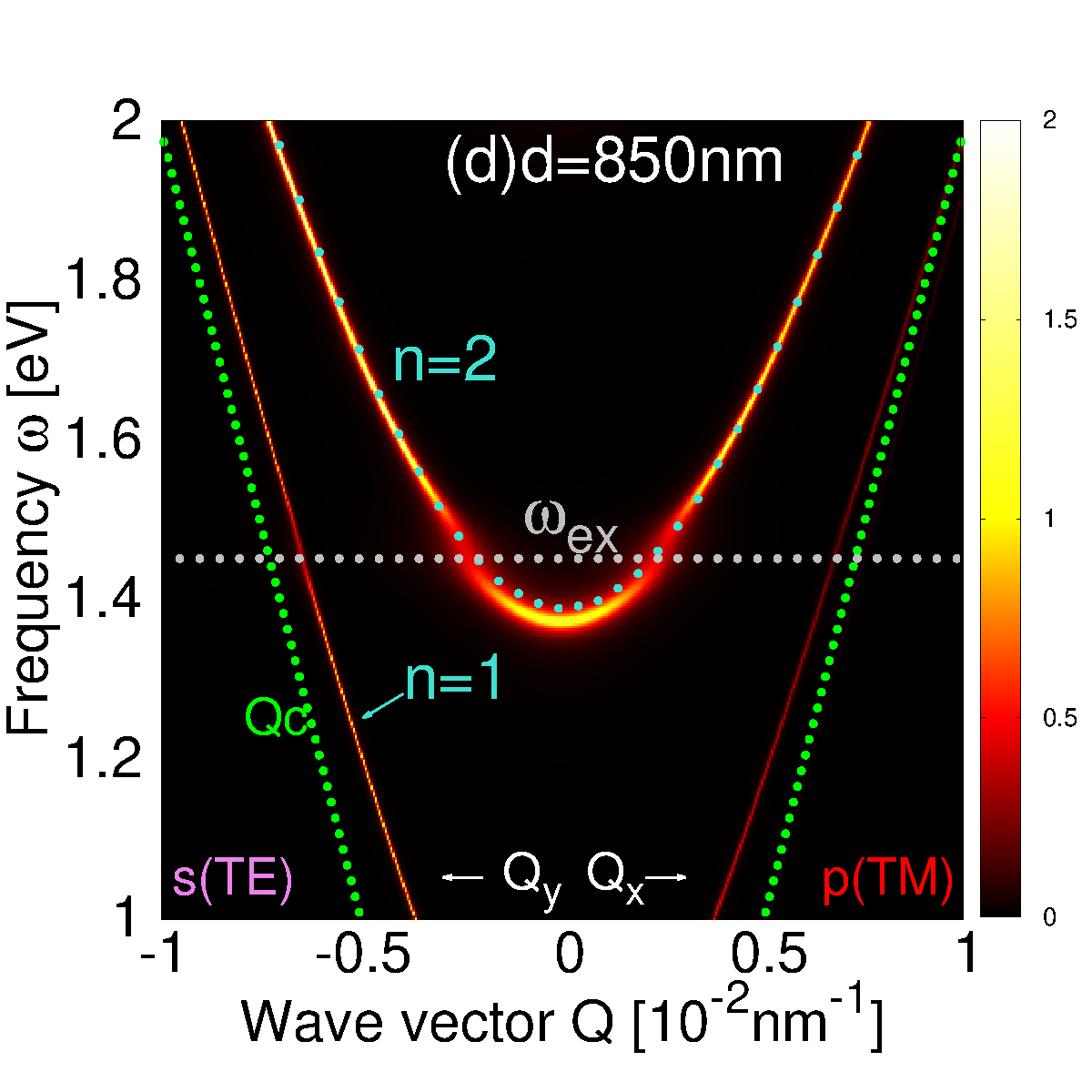}
\caption{The intensities of electromagnatic modes (-Im${\cal E}_{xx}$) showing the hybiridisation 
between silver cavity mode $n=1$ (magenta dotted) and P$_4$ exciton (white dotted) 
for various cavity sizes: (a) $d=375$\,nm, (b) $d=400$\,nm, and (c) $d=425$\,nm, where $z_0=d/2$. (d) The hybridisation between 
cavity mode $n=2$ and P$_4$ exciton in cavity of thickness $d=850$nm and $z_0=d/4$.
}  
\label{Fig3}
\end{figure*}
where $\sigma^{RPA}$ is the RPA optical conductivity\,\cite{Pol,Polariton2016}, while the ladder optical conductivity is $\sigma^{ladd}=\frac{i}{\omega S}j{\cal K}j^*$, where $S$ is a normalization surface and $j$ are the current vertices. The $4$-points polarizability ${\cal K}$ satisfies the Dyson equation\,\cite{sm} in which 
the Bethe-Salpeter-Fock kernel $\Phi^{F}$ enters\,\cite{BSE1,BSE4,BSE9,GWtheory,BSE8}.  
The electron energies  are corrected using the GW method so that the present approach is equivlent 
to time-dependent screened Hartree-Fock approximation\,\cite{BSE1,BSE4,BSE8,GWtheory,BSE9,BSE2}. 
For more details see Secs.\,S1.B and S2.B in Ref.\,\cite{sm}. 
Without cavity photons, the total optical absorption Re\,$\sigma_x(\omega)$ in 
P$_4$ and hBN is characterized with the strong excitons appearing at $\hbar\omega_{ex}=1.45$\,eV and $\hbar\omega_{ex}=5.67$\,eV, respectively. In addition, the absorption spectrum Re\,$\sigma_x(\omega)$ in WS$_2$  shows the two spin-orbit splitted A and B excitons at $\hbar\omega_{ex}^A=2.1$\,eV and $\hbar\omega_{ex}^B=2.46$\,eV\,\cite{sm}.   
\begin{figure}[h]
\includegraphics[width=8cm,height=6cm]{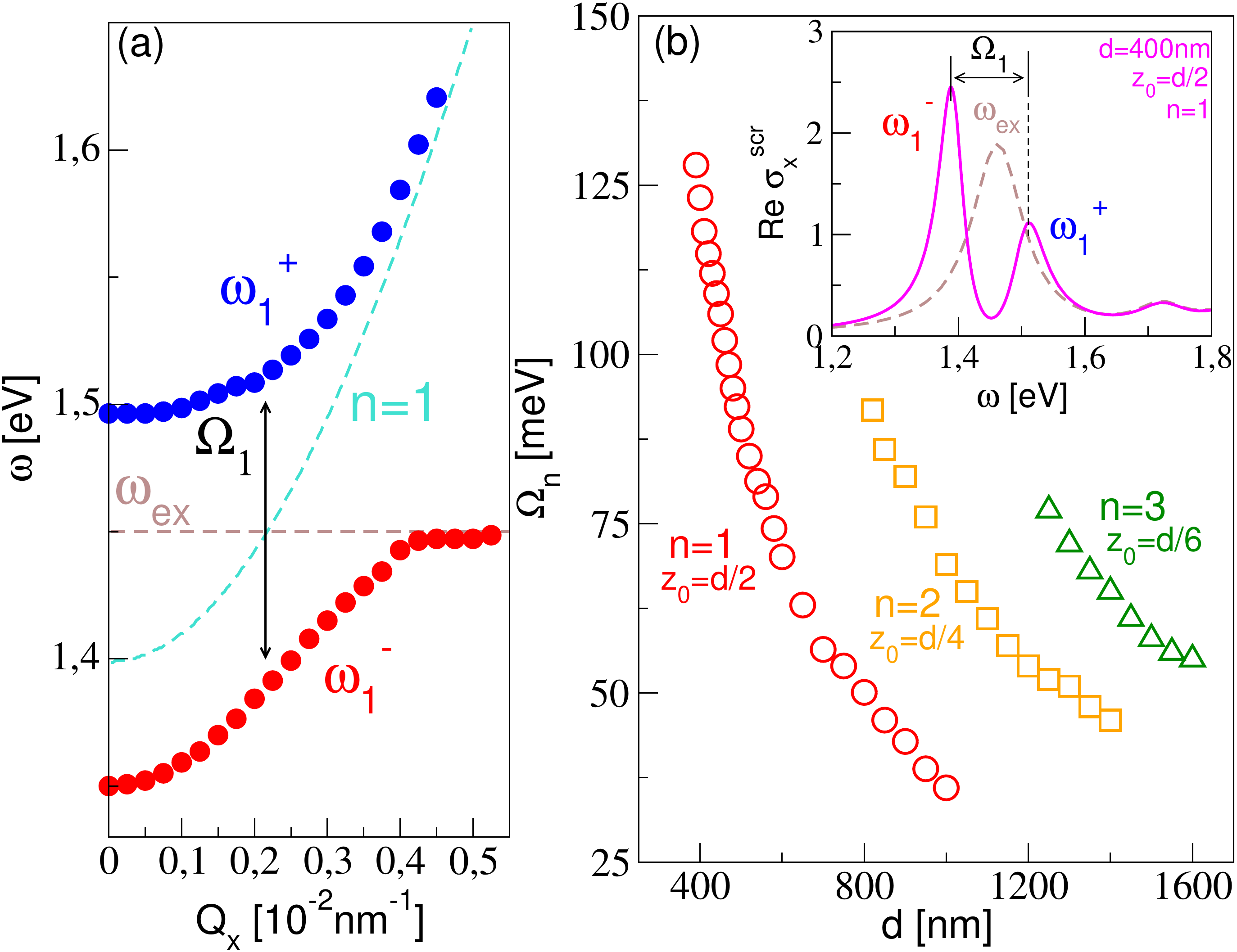}
\caption{(a) The dispersion relations of exciton-polaritons $\omega^-_1$ and $\omega^+_1$ (dots) for P$_4$ in silver microcavity of thickness $d=400$\,nm and $z_0=d/2$. Dashed lines show unperturbed modes. (b) The Rabi splittings $\Omega_n=\omega^+_n-\omega^-_n$ versus cavity 
thickness $d$ for $n=1$ (red circles),  $n=2$ (orange squares)  and $n=3$ (green triangles) cavity modes, where 
$z_0=d/2$, $z_0=d/4$ and $z_0=d/6$, respctively. The inset shows Re\,$\sigma^{scr}_x$ before (brown dashed) and 
after (solid magenta) the P$_4$ is inserted in the cavity, where $d=400$\,nm, $z_0=d/2$ and $Q_x=0.00217$\,nm$^{-1}$ (momentum for 
which exciton crosses the $n=1$  mode).
}  
\label{Fig4}
\end{figure}

Figures \ref{Fig3}(a), \ref{Fig3}(b), and \ref{Fig3}(c) show the modifications of the $n=1$ cavity 
mode intensity after the single-layer P$_4$ is inserted in the middle $z_0$=d/2 ($n=1$ antinodal plane)  of the silver cavity, where the cavity sizes are $d=375$\,nm,  $d=400$\,nm and  $d=425$\,nm, respectively. 
White and turquoise dotted lines denote the P$_4$ exciton and unperturbed cavity 
mode $n=1$, respectively. For $d=375$\,nm, just before the $n=1$ cavity mode crosses the exciton,  a significant part of the $n=1$ mode 
spectral weight is transferred below the exciton energy.  
By increasing the cavity size, i.e. for $d=400$\,nm and  $d=425$\,nm, the exciton crosses the $n=1$ mode, which results in the intensity weakening and band-gap oppening in the intersection area. 
This behaviour enables creation of exciton-polariton condensate, as  experimentaly 
verified in Refs.\,\cite{bec3,ex-po-WSe2,Peroskite_ACSPhotonics,ex-po-perovskite,GaAsQW}. 
By changing the cavity thickness, the exciton can interact also with the higher cavity modes. Figure \ref{Fig3}(d) shows the modifications of $n=2$ mode intensity, where the 
cavity thickness is $d=850$\,nm and P$_4$ is choosen to be located at $z_0=d/4$nm ($n=2$ antinodal plane).  
The exciton significantly weakens the intensity of $n=2$ mode in the intersection area, 
however, here the avoided crossing behaviour is not clearly noticeable in comparison with the exciton coupled to the 1st cavity mode.

The  dispersion relation of exciton-polaritons $\omega^-_n$ and $\omega^+_n$  (hybridised cavity photon-exciton modes), as the one shown in Fig.\,\ref{Fig4}(a), can be precisely determined by following the splitted maxima in 
induced current $j_\mu=\sigma_\mu^{scr}E_\mu$ driven by external (bare) field $E_\mu e^{-i\omega t}$, where the screened optical conductivity is $\sigma^{scr}_\mu=[1-\Gamma\sigma]^{-1}_\mu\sigma_\mu$. 
\begin{figure*}[!t]
\includegraphics[width=4cm,height=4cm]{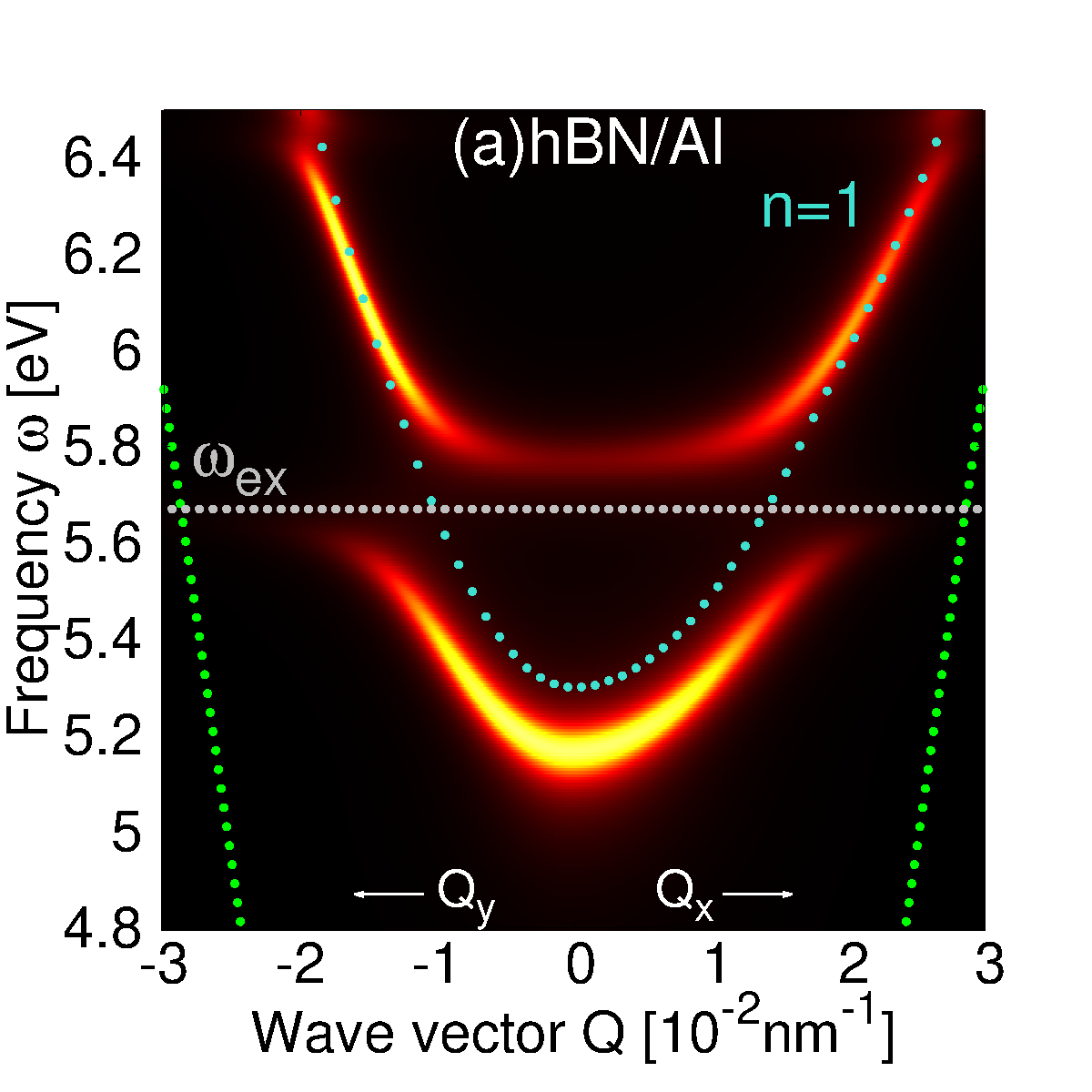}
\includegraphics[width=4cm,height=4cm]{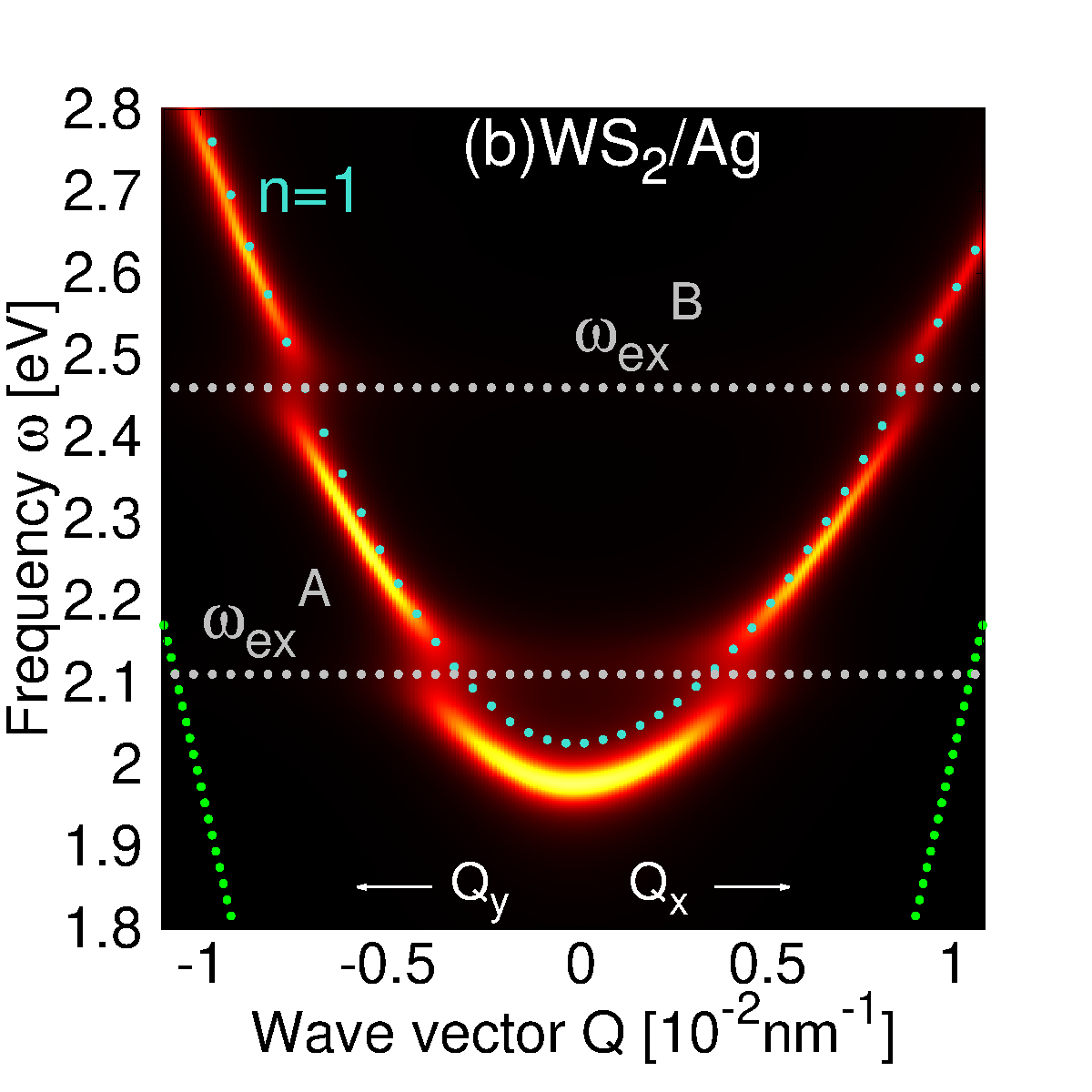}
\includegraphics[width=4cm,height=4cm]{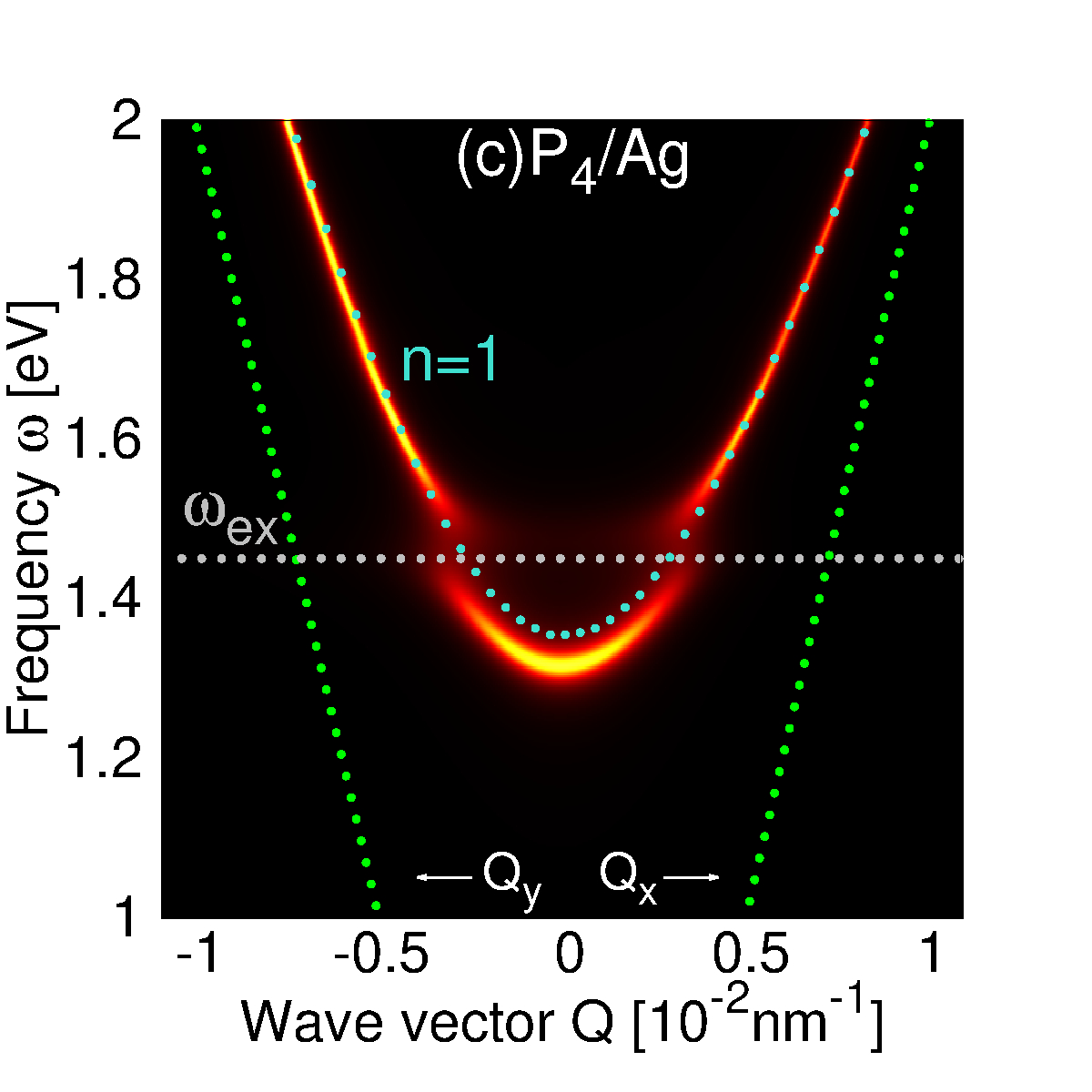}
\includegraphics[width=4.4cm,height=3.69cm]{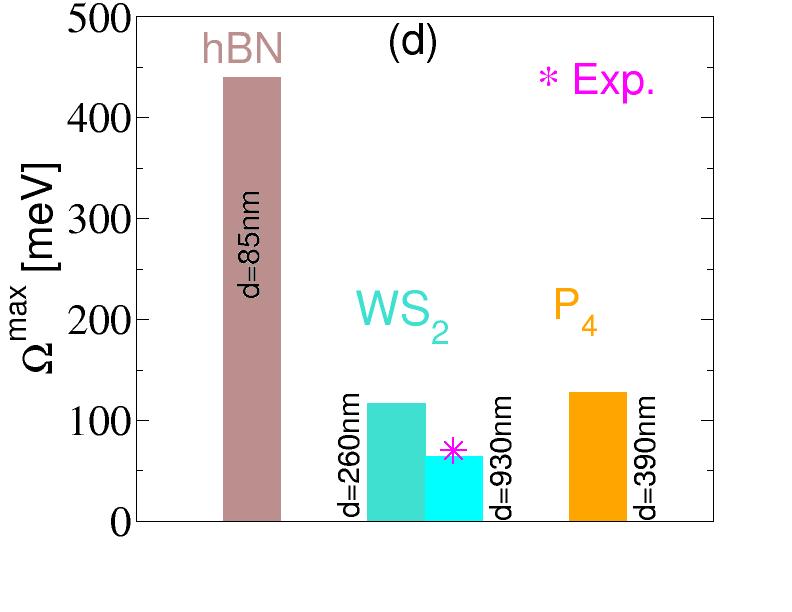}
\caption{The intensity of electromagnatic modes (-Im${\cal E}_{xx}$) showing the 
hybridisation between cavity mode $n=1$ (magenta dotted) and (a) hBN exciton (white dotted) in aluminium cavity of thickness $d=90$\,nm, 
(b) the WS$_2$  A and B excitons in silver cavity of thickness $d=260$\,nm, and (c) the P$_4$ exciton in silver cavity of thickness $d=415$\,nm. The $z_0=d/2$ in all cases. 
(d) Summary of the obtained Rabi splittings in studied systems. The experimental value for WS$_2$ microcavity is from Ref.\,\cite{ex-pol-WS2}.} 
\label{Fig5}
\end{figure*}
The inset of Figure \ref{Fig4}(b) shows the Re$\sigma^{scr}_x$ before (brown dashed) and after (solid magenta) the P$_4$ is inserted in the middle of the cavity of thickness $d=400$\,nm. 
The spliting of exciton $\hbar\omega_{ex}$ to exciton-polaritons 
$\omega^-_1$ and $\omega^+_1$ can be clearly seen. The exciton-photon binding strength can be  
determined from the Rabi splitting defined as difference $\Omega_n=\omega_n^+-\omega_n^-$ for wave 
vector ${\bf Q}$ and for which the bare cavity modes $n=1,2,3,...$ crosses the exciton $\hbar\omega_{ex}$.  
Figure \ref{Fig4}(a) shows the dispersion relations of plasmon-polaritons 
$\omega^-_1$ and $\omega^+_1$  obtained by following the splitted maxima in Re\,$\sigma^{scr}_{x}$ for different wave vectors $Q_x$, $d=400$\,nm and $z_0=d/2$. The clear anticrossing behaviour and Rabi splitting of $\Omega_1=123$\,meV indicates strong interaction between exciton and $n=1$ cavity photon. Red circles, yellow squares and green triangles in Fig.\,\ref{Fig4}(b) show the Rabi splittings $\Omega_n$ for $n=1$, $n=2$ and $n=3$, respectively, versus cavity thickness $d$. The maximum Rabi splittings of $n$th mode $\Omega^{\rm max}_n$ are achieved when $z_0=d/2$ and for $d$ choosen so that $n$th mode just 
starts to cross the exciton energy $\hbar\omega_{ex}$. All three modes show strong coupling with exciton that results in the maximum splitings of $\Omega^{\rm max}_1=128$\,meV,  $\Omega^{\rm max}_2=92$\,meV  
and $\Omega^{\rm max}_3=77$\,meV for $d=390$\,nm, $d=820$\,nm and $d=1250$\,nm, respectively. For larger $d$ the cavity modes cross exciton at a greater angle and the splitting decreases. 
A decrease of $\Omega^{\rm max}_n$ as $n$ increases confirms a confinement hypothesis; as $n$ increases the cavity photon modes crosses the exciton 
for larger $d$, and photon becomes less confined while the coupling is reduced. Thus, the coupling will be stronger as the thickness 
$d$ at which the crossing between $n$th mode and exciton occurs is getting smaller.

The above criterion is met by excitons with higher excitation energy, such as for instance the UV exciton in the hBN single layer. 
Since in the same UV frequency region the cavity should be highly reflective (i.e., $\omega_p>\omega_{ex}$), the appropraite cavity for hBN layer can be made of aluminium with $\omega_p\approx 15$\,eV. Figure \ref{Fig5}(a) shows the modification of aluminium $n=1$ cavity 
mode intensity after the hBN monolayer is inserted in the middle of cavity of thicknesses $d=90$\,nm. The dotted lines show the unperturbed cavity 
mode $n=1$ and the hBN exciton. The strong exciton-photon intercation results in 
strong modification of the $n=1$ cavity mode, band-gap opening and Rabi splitting of 
$\Omega_1=400$\,meV. The maximum Rabi splitting of $\Omega^{\rm max}_1=440$\,meV  is achieved for 
$d=85$\,nm. This large excion-polariton band gap and Rabi splitting suggest a 
possibility of experimental realization of robust exciton-polariton condensate in the cavity setup.

The 2D exciton-polaritons are experimentally studied mostly 
in various TMDs where the mesured Rabi splittings of $\Omega=46$\,meV, $\Omega=26$\,meV, and $\Omega=20$\,meV are found in MoS$_2$\,\cite{ex-pol1}, WSe$_2$\,\cite{ex-pol2},  and MoSe$_2$\,\cite{ex-pol22}. For WS$_2$ the experimentally measured splittings are around $20-70$\,meV for $d>1\,\mu m$, depending on the precise cavity size\,\cite{ex-pol-WS2}. In Fig.\,\ref{Fig5}(b) we show the modification of the silver $n=1$ 
cavity mode intensity when the WS$_2$ monolayer is inserted in the middle of microcavity of 
thicknesse $d=260$\,nm. The unperturbed $n=1$ mode as well as the A and B excitons of bare WS$_2$ are also denoted by dotted lines. 
Both excitons significantly perturbe the $n=1$ mode providing  
the Rabi splitings of $\Omega_1^A=117$\,meV and $\Omega_1^B=103$\,meV.
For comparison, Fig.\,\ref{Fig5}(c) shows 
the modification of $n=1$ mode intensity when P$_4$ is inserted in silver microcavity for the same 
conditions as in WS$_2$ microcavity presented in Fig.\,\ref{Fig5}(b); $n=1$ minimum is $100$\,meV below the exciton. 
The cavity thickness is $d=415$\,nm and $z_0=d/2$. Interestingly, the achieved Rabi splittig is here 
also $\Omega_1=117$\,meV, even though according to confinement hypotesis, the A exciton, which is confined in smaller cavity, 
is expected to split more. However, the A exciton in WS$_2$ has 
smaller oscillatory strength than P$_4$ exciton [cf. Figs.\,S4(a) and S5(b) in Ref.\,\cite{sm}] so that the binding is weaker and the two effects cancel.

Finally, in Fig.\,\ref{Fig5}(d) we compare the maximum splittings of exciton-polaritons $\Omega^{\rm max}$ for the three semiconducting microcavities, summarizing the different regimes of exciton-cavity photon coupling strengths in these materials. For WS$_2$ microcavity we additionaly present the results for the experimentally measured value of cavity size, i.e., $d=930$\,nm\,\cite{ex-pol-WS2}, when exciton interacts with $n=2$ cavity mode. The corresponding value of $\Omega^A_2=64$\,meV shows an excellent agreement with the experiment\,\cite{ex-pol-WS2}. For the same cavity thickness, splitting of the $n=1$ cavity mode and the WS$_2$ A exciton is $\Omega_1^A=42$\,meV.

In summary, we have studied the intercation strengths between cavity photons and excitons in various 2D semiconducting crystals by means of rigorous \emph{ab initio} methodology. It is shown that insertion of 2D crystals into a metallic microcavity significantly modifies the photon dispersion. For instance, the band gap opening and Rabi splitting as high as $\Omega=440$\,meV was obtained for hBN cavity device. This opens a possibility of experimental realization of the robust 2D exciton-polariton condensate. Moreover, the exciton-photon interaction strongly depends on photon confinement, which was shown to be adjustable by the cavity thickness $d$. The results of exciton-polariton splitting in WS$_2$ cavity device show a good agreement with recent experiments and suggest higher photon confinements with decreasing cavity size at which stronger photon-matter coupling should be achieved. In order reach this stronger binding we suggest an experimental setup consisting of tunable submicrometer cavity (such as AFM-tip and substrate) tuned so that the principal photon cavity mode coincide with the exciton energy, e.g. as in Fig.\ref{Fig4}(b).

\begin{acknowledgements}
The authors acknowledge financial support from European Regional Development Fund for the ``QuantiXLie Centre of Excellence'' (Grant KK.01.1.1.01.0004) and ``Center of Excellence for Advanced Materials and Sensing Devices'' (Grant No. KK.01.1.1.01.0001), as well as from Croatian Science Foundation (Grant no. IP-2020-02-5556 and Grant No. UIP-2019-04-6869). Computational resources were provided by the Donostia International Physic Center (DIPC) computing center.
\end{acknowledgements}

\end{document}